\documentclass[12pt]{article}
\begin{document}
\title{{\bf Is Nature Generic?}}
\author{Spiros Cotsakis and  Peter Leach\thanks{Permanent address: School of Mathematical
and Statistical Sciences, University of Natal, Durban 4041,
Republic of South Africa}\\
\\ GEODYSYC\\ Department of
Mathematics
\\ University of the Aegean\\ Karlovassi 83 200, Greece}
\maketitle
\begin{abstract}
\noindent An introductory guide to mathematical cosmology is given
focusing on the issue of the genericity of various important
results which have been obtained  during the last thirty or so
years. Some of the unsolved problems  along with certain new and
potentially powerful methods which may be used for future progress
are also given from a unified perspective.
\end{abstract}

\section{Introduction}
We live in space and time. For the cosmologist this fact relates
to some fundamental and unresolved issues:
\begin{itemize}
\item  How was our spacetime created?

\item  What is the shape of our space? Was it always the same? What are the
possible `admissible' shapes for our physical space?

\item  Was our spacetime so `simple' in the past or more complex? What about
in the future?

\item What was the structure of the `early universe'?

\end{itemize}

\noindent and so on. These issues, when translated into a suitable
mathematical language, do in fact drive most of the current
research in mathematical and theoretical cosmology.

In this paper we lay the foundations of modern mathematical
cosmology in a manner suitable for the nonspecialist or a graduate
student who wishes to have some initial orientation in his/her
attempts to embark on research in this fascinating field of
Science which lies in the interface between Applied Mathematics
and Theoretical Physics. We have tried to present (an outline of
the elements of) mathematical cosmology from a very broad
perspective suitable for many readers and hope that even experts
who work in one or more of the many modern branches of cosmology
will find here some points of interest.

In the next Section we present the basic principles of
cosmological modelling. Modelling the universe presents some new
challenges for the applied mathematician or theoretical physicist
different from those in other areas of the mathematical modelling
of physical phenomena. Section 3 introduces and discusses the
fundamental notion of {\em a cosmology.} Section 4 gives an
overview of the basic unsolved problems of mathematical cosmology
and the broad lines of attack that have been and are still being
used by different research groups as well as some new and
potentially efficient mathematical methods which could powerfully
augment the successful treatment of the cosmological problem.
Conclusions and future prospects are given in Section 5. Although
almost no references are given in the text, the Bibliography
presents some very basic items which are meant to serve as a
useful entrance to the literature of this vast and truly exciting
subject.

\section{Principles of cosmological modelling}
Modelling the universe, as opposed to that of other physical
systems, is an involved and unique process different in nature and
scope from other modelling in mathematical physics. There
are two basic steps in the process, one  we may call the
theoretical step (items 1-3 below) and secondly the observational
step (item 4 below). These two steps  comprise in turn 3+1 basic
features:
\begin{enumerate}
\item  A (cosmological) spacetime

\item  A theory of gravity

\item  A collection of matter fields

\item  The process of confronting the results of suitable combination(s) and
analyses of 1-3 with the unique observed universe
\end{enumerate}
We can loosely define {\em a cosmology} as the result of the
appropriate combination of the features 1-4 above. The unifying
principle that ties the basic features 1-3 together to form what
we call {\em a cosmological model} is the {\em Action Principle}.
Let us first consider in some detail the three most basic
constituents of a cosmology.

\subsection{Spacetimes}
There is a basic cosmological hierarchy of spacetimes according to
the degree of exact symmetry involved.
\begin{itemize}
\item  Isotropic (Friedmann-Robertson-Walker) spacetimes

\item  Homogeneous (Bianchi) nontilted spacetimes

\item  Homogeneous (Bianchi) tilted spacetimes

\item  Inhomogeneous $G_{2}$ spacetimes

\item  Inhomogeneous $G_{1}$ spacetimes

\item  Generic spacetimes
\end{itemize}
Bianchi is that family  of homogeneous but anisotropic spacetimes
first classified by the Italian geometer L. Bianchi according to
 the underlined Lie algebra in nine types $I,\dots IX$ and two
 classes A, B. This is the most general family of spacetimes for
 which the Einstein field equations reduced to ordinary differential equations
 since the space dependence of the metric derivatives are suppressed.
 Here, $G_{i}, i=1,2$ means the group of symmetries ($i$ indicates  the number of
Killing vectors) of the underline spacetime manifold.  The group $G_{2}$ is
larger than $G_{1}$ and consequently the spacetimes in the category four above
are more symmetric than those in category five. It can be shown that, in a
certain sense which can be made
precise, spacetimes in a given category above are contained in the
next category as special cases. Therefore we have a list of
increasing generality (top to bottom) or genericity and the first
families in the list may not be considered as realistic candidates
for the actual universe as they all contain (or are constructed
through the use of)  {\em exact} symmetries. However, they are
very important as toy models as well as simpler cases which may
contain the seed of the true dynamics of the more generic (but
essentially more difficult to handle mathematically) spaces. The
ones without any symmetry are in the last category, generic
spacetimes, while those with maximal symmetry  are the isotropic
spaces. The latter are the most common spacetimes used in
cosmology today.

The first five families of spacetimes are basically formed by
having a group of transformations {\em acting} in some way on the
spacetime manifold such that its orbits essentially create the underline
point set (the action of the group on the manifold is then called
{\em transitive}). The dimension of this group as well as the
manner it acts on the manifold are responsible for the wide
variety of cosmological spacetimes. The simplest ones are the
isotropic spacetimes whereas generic spaces are extremely
difficult to handle.

\subsection{Theories of gravity}
Since the realization that, under certain assumptions, general
relativity leads to singularities and consequently may not
correctly or adequately describe the "observed" features of the
universe at very small distances or very high energies, there has
been an endless process of constructing new theories which
incorporate gravity but extend general relativity in many
different ways. Here is an incomplete list:
\begin{enumerate}
\item  General relativity (GR)

\item  Higher derivative gravity theories (HDG)

\item  Scalar-tensor theories (ST)

\item  Supergravity theory

\item  String theories

\item  Branes

\item  M-theory

\item  $\cdot \cdot \cdot $
\end{enumerate}
As we shall see, one of the basic problems in cosmology is how to
figure out which theory of gravity may be the most suitable for
describing the universe at its early stages of evolution. Many
cosmologists believe that perhaps some variant of string theory
must be the final word, but choosing a gravity theory for such a
purpose certainly involves many different and interrelated issues.
We consider this problem in more detail below.

The basic method used to construct and compare all these different
theories is the Action Principle, familiar from Classical
Mechanics we learn as students. This principle forms
the basis of modern Theoretical Physics and in fact, all the
gravity theories above come out by postulating the Action
Principle. HDG theories extend GR by the addition of extra terms
in the gravitational action functional (a function defined on the
space of metrics), terms which contain higher powers of the
curvature invariants. ST theories postulate that the gravitational
field is mediated by a scalar field in addition to the spacetime
metric, the simplest  prototype of this family being the
well-known Brans-Dicke theory. This class of gravity theories is a
very broad one incorporating in effect many of the string theories
as special cases. There has been known for some time that there
are certain conformal `dualities' between HDG, ST theories and GR
in that these theories are GR in disguise with additional
`fields'. Dualities have also been recently discovered between
different string and supergravity theories and these in turn
sometimes are interpreted to imply the existence of a more general
theory,  which might in some subtle way incorporate all previous
ones as special cases, M-theory.
\subsection{Matter fields}
Here too one may easily compose a shopping list of interesting
candidates for matter fields which may have played an important role
during different epochs in the history of the universe.
\begin{itemize}
\item  Vacuum

\item  Fluids

\item  Scalar fields

\item  Wave maps

\item  Electromagnetic fields

\item  Yang-Mills fields

\item  $n-$form fields

\item  Spinors
\end{itemize}
Each one of these different families has its own special role to play
in cosmology, but some are definitely more ambiguous than others
for different reasons.

With this background, let us now see how modern cosmologists put together
spacetimes, gravity theories and matter fields to form the basic ingredient
of their subject, {\em a cosmology.}

\section{Cosmologies}
How do we construct a cosmology? Pick up a spacetime from the
cosmological hierarchy list, choose a gravity theory and one or
more matter fields, tie them together through the Action Principle
and try to explain the observed facts in terms of the consequences
 of the application of the variational principle. The result is called {\em a
cosmology. }In the form of a symbolic equality,

{\bf Cosmology = Cosmological model(s) + Observations}

We shall denote a given  family of cosmological models (or a
cosmology) with a triplet $\{\cdot /\cdot /\cdot \}$ of the sort
\{Spacetime/Gravity theory/Matter field\}. The simplest and best
studied (relativistic) cosmology of physical interest is the
\{FRW/General Relativity/Fluid\} Cosmology. This actually  is the
cosmology discussed in many textbooks on the subject under the
heading `Relativistic Cosmology'.

One may obviously attempt to construct and analyse other
cosmologies, based for example on the families:
\begin{itemize}
\item  FRW/GR/vacuum cosmologies

\item  FRW/HDG/vacuum cosmologies

\item  FRW/scalar-tensor/fluid cosmologies

\item  FRW/string/vacuum cosmologies

\item  FRW/Brane/scalar cosmologies

\item  Bianchi/GR/fluid cosmologies

\item  Bianchi/GR/scalar cosmologies

\item  Bianchi/scalar-tensor/vacuum cosmologies

\item  Bianchi/M-theory cosmologies
\end{itemize}
\noindent and so on.  How do we study the properties of each one of
these cosmologies? There are many questions we can ask, some common
to all families and others  particular to some family. We
have more to say on this in the next Section. The important
thing is that the families we construct be mathematically
consistent, toy models upon which to base our physical predictions
and  conclusions for the structure of the universe at different
epochs in cosmic history.

A particular issue, connected with the philosophy that there is
not one single theory of the universe which would describe it at
all times but some cosmologies may be more adapted to some epochs
while others not, is the {\em problem of cosmological cohesion},
that is to try to connect different cosmological models together
to form a consistent frame, {\em a cohesive cosmology},  to
compare with observations. For example, suppose that an
FRW/GR/fluid cosmology is valid  after the Planck time
onwards and that a Bianchi/M-theory/vacuum cosmology holds well
before that time. The cosmological cohesion problem in this case
is  to connect the physically meaningful solutions of the
two cosmology branches into one cohesive cosmology that would
describe the entire cosmic history and be compatible with
observations and other constraints.

The cohesion problem is one between {\em different} cosmologies
which have already been studied and their solution spaces are more
or less clarified. However, the first step in the study of
cosmologies is to single out a particular family and to try to
develop a well-defined theory addressing as many issues as one can
in the garden of cosmological problems. Some of these are
described in the next Section.

\section{Cosmological problems}
We now translate the questions stated in the Introduction in a
more suitable terminology. The result is a number of very broad
directions of research currently pursued.

\subsection{The singularity problem} This is the ultimate and
most important problem that every cosmology has to face. It
indicates the true range of validity of any cosmology and of
course that of the underlying gravity theory. Its two parts,
namely, the existence and the structure/nature of singularities
are very different and may play complementary roles in deciding
the final fate of any theory of cosmology.

Usually in cosmology the definition of a singularity is taken to
mean a place where some physical quantities, for example the
spacetime curvature, densities, temperatures of matter fields etc,
become {\em infinite} or discontinuous there. Hence it is usual
that cosmological singularities are connected with either
infinities or  pole like behaviour. As such, it is not surprising
that there is little to be said about their structure or nature
using the usual geometric/topological methods. Indeed the
singularity theorems in general relativity  are geometric
existence results about incomplete geodesics the endpoints of
which, strikingly, coincide with infinite curvature singularities
in most cases but the nature of these singularities is undecided
in the general case.

Instead the nature of singularities is commonly tackled via the
methods of dynamical systems for particular cosmology families. In
the first three families of the basic spacetime hierarchy, we end
up typically with ordinary differential equations whereas from the
last three categories we find systems of partial differential
equations. Using methods borrowed from the qualitative theory of
differential equations (theory of dynamical systems) cosmologists
have been able to figure out the behaviour of spacetime in the
vicinity of a singularity. In the most general case wherein the
Einstein equations are reduced to ordinary differential equations,
that is the second and third families in the hierarchy, very
complex structures can appear in the neighborhood a such singular
points. A basic question is whether such structures remain as
generic features in the more general cases down the hierarchy or
disappear when we consider more general cosmologies as a result of
the less and less symmetry imposed.

An special example of the singularity issue in cosmology is {\em
the recollapse problem}, that is whether or not all closed
(compact, without boundary) cosmologies recollapse to a second
singularity. Of course, in general it is very easy to construct
examples where closed universes filled with special matter fields
do not  recollapse,  but the question here is, given a cosmology,
under what conditions does the subclass of all closed cosmologies
recollapse to a singularity in the future. This purely classical
problem acquires importance also in the framework of inflationary
and quantum cosmology since it is yet to be decided whether or not
the universe can recollapse before an inflationary phase is
reached (the so-called premature recollapse).

\subsection{The problem of cosmic topology.} This problem has two
aspects. The observational problem of deciding what the shape of
the observed universe is and what would be the consequences of
supposing that space has a different topology that the usual one.
For example, there are many known examples of different topologies
(euclidean, torus)  of which  can all admit a flat metric. The
supposition that the manifold geometry is hyperbolic has become
very fashionable and attracts a lot of attention currently. As W.
Thurston has put it, `it is a wonderful dream to see the topology
of the universe some day'. Perhaps the topology of the universe is
non-trivial but not very complicated.

 The second,
theoretical, aspect is more involved with apparently many
consequences for different parts of the general cosmological
problem, most of which are as yet unclear. It is well-known that,
although the Einstein equations evolve only the geometry (that is
the spacetime metric) but leave the topology of  initial data sets
fixed (but arbitrary) during the evolution there are cases, for
instance the formation of singularities in the future, where the
topology of the initial data set which  evolves is expected to be
different after some of the space has collapsed. In fact, this
issue seems to be related in a subtle way to the fundamental
problem of classifying 3-manifolds (in this case the initial data
sets). The Einstein flow evolves such data and one would like to
know how initially different topologies affect the flow and vice
versa.

\subsection{ The problem of asymptotic states.}

The existence or nonexistence of singularities in a particular
cosmology, notwithstanding, the issue of providing a detailed
description of the dynamical behaviour of cosmological spacetimes
at both small and large times in a particular gravity theory with
matter fields is a very important one, aiming at establishing a
first test of the scope and flavor of any particular candidate
cosmology.

One may tackle this problem by finding particular exact solutions
that describe special families of models within a given cosmology
and this has been in fact the first line of attack in modern
cosmological research. However, even if one has succeeded in
finding many different solutions of a particular cosmology at hand
it is often difficult to combine them into a coherent whole that
would indicate the true picture of dynamical possibilities of the
given family.

The dynamical systems approach used by several authors in the past
is much more promising when we wish to uncover the global
structure of the solution space of a given cosmology. An
especially important example of an aymptotic problem is the
so-called {\em the isotropization problem} (or in other contexts
{\em a cosmic no-hair conjecture}) which aims at examining the
possibility of first accepting that at an early stage in cosmic
history the universe was in some more complex state described by
one of the models down the hierarchy list (eg., Bianchi) and then
showing how the present isotropic state is the result of the long
term, `observed', dynamical evolution of that less symmetric
`initial' era. This is where the central dynamical concepts of
trapping sets and attractors may come into full play. What are the
attractors of a given cosmology? Is it possible that attractors of
one cosmology are related to attractors of another? An answer to
this question will clarify to what extend members of one cosmology
family belong also to another family and in this sense how
different cosmologies are related to one another. A {\em global
attractor} of a cosmology is defined as one that attracts all
neighboring members inside the given cosmology. If one could show
how the global attractors of {\em different} cosmologies are
related, one would have a precise way of deciding which cosmology
to pick for specific eras in the cosmic history. Consequently such
a result would help to connect apparently different cosmologies.

Another issue is to find whether {\em chaotic, unpredictable
dynamical behaviour} is a true feature of classical cosmological
dynamics. The basic dichotomy of nonlinear dynamics, namely,
integrability versus nonintegrability and chaos, is certainly to
be found in mathematical cosmology too.  Only the simplest
cosmologies are translated into two dimensional dynamical systems
and most of them are of dimension higher than four. Therefore
complex dynamical behaviour is generally to be expected in
cosmology and indeed this has been a subject of considerable
research in modern mathematical cosmology. The notion of a
cosmological attractor introduced earlier will also play a special
role here as dissipative cosmologies are generally expected to
have the so-called {\em strange} attractors, but their existence
may not be easy to unravel except in the case of highly symmetric
cosmologies.

An emerging method in recent years to decide whether a given
cosmology is integrable, without actually solving the associated
differential equations to construct cosmological solutions of
physical interest, is based on an intriquing idea of two great
mathematicians of the past, S. Kowalevski and P. Painlev\'{e}.
These people thought that instead of trying to solve the relevant
differential equations which describe a given problem, it would be
very convenient to decide whether any given system (hamiltonian or
not) is integrable if there was a way to merely examine  the {\em
form} of it. The answer appears to lie in the complex plane and
the types of singularities the equations can have when
analytically continued in the complex time plane. Kowalevski was
able to discover a new, as well as recover many of the integrable
cases of the so-called Euler-Poisson equations that are associated
with the problem of the Lagrange top by analysing these equations
when {\em the only movable singularities that the equations can
exhibit in the complex plane are poles.} This feature is called
the Painlev\'{e} property and  has become very important in recent
years in attacking the integrability problem. If this holds, then
all solutions lie in a single Riemann sheet. However, much more
complicated behaviour can occur when the singularities of the
analytically continued system fail to be poles but take the form
of {\em movable branch points} or the even {\em essential
singlarities} for which the solutions are in general multivalued
complex functions. The general {\em integrability conjecture} is
that if a system has the Painlev\'{e} property then it is
integrable. Although, not completely proven, this method has been
applied with great success in many systems in Mathematical Physics
in general and  in mathematical cosmology in particular. The
reasons why such a method seems apparently to work appears to be
connected with algebraic geometry and the theory of elliptic
curves. One is therefore hoping that the complicated behaviour
seen in the vicinity of the big-bang singularities in many
cosmologies could be quantified by using this method and looking
at the singularity patterns which the analytically continued
solutions of the real-time systems form on approach to the
cosmological singularity. This program is still at an infant
stage, but it is has the potential to yield interesting results in
the coming years.

\subsection{ Gravity theories and the early universe. }
This last problem is a very difficult and basic one to all
attempts to construct a realistic, cohesive cosmology. It is
evident that the issue of choosing a gravity theory with which to
build a cosmology is of paramount importance to cosmological model
building. It has been known for many years and  regarded as
folklore that general relativity cannot be meaningfully
extrapolated back to very early times in the history of an
expanding universe which is predicted by it. A look at  the list
of possible alternative gravity theories, however, reveals that
none is thus far the unique, problem-free theory. Each time there
is some particular theory which is fashionable, M-theory being
today's choice.

One way to decide among a host of possibilities has been to try to
get a feeling for how these different cosmologies behave when we
ask the same questions.  This leads to a picture of dynamical
possibilities for the whole set of all conceivable cosmologies and
 indeed, has been the Holy Grail of modern research in
 mathematical cosmology. The  picture to-date, however, is by no means complete
even in General Relativity and the search is continued. Here again
we see the need for the full exploitation of the cosmological
attractors in an effort to understand the precise relations
between different cosmologies.

Although a recurrent theme in this paper has been the fact that we
should work among all different possibilities (and this in fact is
a basic characteristic in current research in the field), one may
think that, for the case where General Relativity is expected to
break down, some principle exists that could successfully guide
our vision in  searching for the `right' theory with which to
build a reliable cosmology of the early universe. A principle
suitable for such a purpose can be based on the use of the
fundamental notion of {\em symmetry}. Which cosmology is the most
symmetric? This question raises another: What is meant by "most
symmetric"? The notion of a {\em Lie  symmetry} is a natural one
when applied to systems of differential equations. It is related
to the fundamental invariant quantities that are preserved during
the evolution of the system according to Noether's theorem. If we
could construct the symmetry atlas of any given cosmology we would
have gone a long way to answering any given question about the
evolution of a cosmology. Such an approach is not difficult to
implement and could be done for a great variety of cosmologies.
This may prove to be an interesting and fruitful direction of
further research in early universe mathematical cosmology in the
coming years.

\section{Outlook}
The strategy is now clear. Study each one of these problems in the
framework of every possible cosmology with an effort to decide
among the different possibilities for a realistic cosmology.

 The cosmological problem is the global problem {\em par excellence.}
In contradistinction with the other, complementary area of modern
research in gravitation, namely, asymptotically flat problems, all
basic problems in cosmology involve thinking about spacetimes
which are nowhere trivial and in this sense the lack of knowledge
of initial conditions in cosmology is natural (if only trivial
boundary or initial conditions are acceptable!). Asymptotically
flat problems on the other hand, being basically local ones, are
well-defined mathematically having initial or boundary conditions
away from the sources where the spacetime is trivial. But the
universe is not asymptotically flat. The Newtonian universe is
asymptotically flat, but general relativity introduced the notion
of an evolving universe as a whole and therefore did away with
asymptotic flatness on a cosmological scale.

The fundamental problems of mathematical cosmology discussed in
this paper, namely the singularity problem, the topology problem,
the asymptotic problem and the problem of choosing a gravity
theory and building a realistic early universe cosmology, frame
mathematical cosmology as a separate and important discipline at
the interface between Mathematics and Physics and make it an
interesting and active branch of Mathematical Physics.

A new direction of research in the singularity problem might
consist in using the highly developed theory of singularities of
differentiable mappings by Arnol'd and coworkers. Since the usual
singularity theorems prove the existence of  families of
incomplete geodesic curves which in general refocus to form
caustics, perhaps  a clarification of the nature of these
singularities can be attained by their classification through
Arnol'd theory. However, the latter is concerned with
singularities of a different type namely, those that occur due to
the vanishing of certain derivatives and Jacobians rather than
infinities or poles. Is it possible that the singularities in
general relativity be of the milder type of this sort? The answer
to this question is at present unknown.

Much work has been undertaken during the last thirty years or so
in the asymptotic problem using the qualitative theory of
differential equations. This work can be generalized in at least
two directions. Firstly, most of the analyses are concerned with
equilibrium solutions and their stability. Bifurcation theory may
open the way to tackle seemingly unrelated systems as a whole
system with parameters, for example, the general Bianchi/GR/vacuum
family.

Secondly, the equations of the family: $G_{2}$/GR/vacuum are very
similar to those of a spherically symmetric wave map and existence
and regularity results for the latter system are known in the
literature. The theory of partial differential equations theory
has not been used in any systematic way up till now in
Mathematical Cosmology. Global information about the solution
spaces of some of these inhomogeneous cosmologies may also be
obtained by writing them as dynamical systems in infinite
dimensions and some work along this lines is now beginning to
emerge.

Most of the published literature in the early universe cosmology
is mainly concerned with the first two of the six-step spacetime
hierarchy given above. It is entirely unknown to what degree
important discoveries (a prime example is inflation) that have
been made working with `low-level' (ie., top of the hierarchy)
cosmologies (e.g., FRW/GR,HDG,ST,String,Brane/scalar etc) are
justified, i.e., persist as true features of the generic dynamics
or are simply artifacts of the high degree of exact symmetry
imposed. That is an additional reason why the issue of
cosmological attractors in given cosmologies must be faced.

A work that analysed the sixth stage (generic spacetimes) but in
the asymptotically flat case in General Relativity is the proof of
the global stability of Minkowski space by Christodoulou and
Klainerman (Annals of Mathematics Studies, vlm. 41, Princeton
University Press, 1993). No result of such generality exists for
any cosmology. What could a corresponding analysis in the
cosmological case imply (for instance the global stability of the
positive curvature FRW spacetime) for the validity of the current
cosmological ideas (inflation, attractor properties of the known
physically interesting cosmological spacetimes etc) is at present
only a matter of conjecture.

Nature is unique, it is not generic. Our attempts to simulate the
universe in mathematical and theoretical cosmology will lead to
reliable results if and only if they follow from studies of
generic cosmologies or show which features of the highly symmetric
(and hence unphysical) cosmological models persist and propagate
down the hierarchy list so as to become true features of the  more
general, asymmetric cosmologies. Progress will be made if one
finds a way to sidestep the difficulties of analysing the partial
differential equations of the inhomogeneous models by showing how
the global attractors of different cosmologies are related and
picturing more clearly the generic structures of the cosmological
phase space. It is only in this way that the observations showing
a homogeneous universe can be justified mathematically and give
meaning to our ability to work with models high in the
cosmological hierarchy list. On the other hand, if the unique
features of Nature cannot be recovered by a sort of generic
process the road to understanding will be very long and arduous.
\\ \\
{\bf Acknowledgements}
\\

\noindent We thank John Barrow, Yvonne Choquet-Bruhat, George
Flessas and John Miritzis for many useful discussions and kind
comments.

\end{document}